\documentclass[preprint,showpacs,preprintnumbers,amsmath,amssymb,superscriptaddress,aps,pra]{revtex4}
\usepackage{amssymb,amsmath,graphics,epsfig,amssymb,rotating}
\usepackage{color}

\begin{document}

\title{{\bf Two-dimensional three-body quadrupole-quadrupole interactions}}

\author {Jianing Han \footnote[1]{Email address: jhan@southalabama.edu}}
\affiliation{Department of Physics, University of South Alabama, Mobile, Alabama 36688.}

\date{\today}
\begin{abstract}
Similar to interactions between dipoles, or van der Waals interactions, quadrupole-quadrupole interactions are interactions between quadrupoles. In this article, we study the quadrupole interactions between highly excited atoms or Rydberg atoms. In addition, unlike many other calculations, in which the primary focus was on the one-dimensional two-body quadrupole-quadrupole interactions, the primary aim of this article is to study the two-dimensional few-body quadrupole-quadrupole interactions. Specifically, the two-dimensional three-body interactions are investigated. This research has many applications, such as quadrupole-blockade for quantum computing, creating molecules based on quadrupole interactions. 

\label{abst} 
\end{abstract}

\maketitle
\section{Introduction}
Quadrupole-quadrupole interactions have been studied in different areas of research, such as nuclear physics, ionic or plasma physics, and molecular physics. For example, quadrupole interactions between molecular rotation and a spinning nucleus have been investigated \cite{Fano}. Quadrupole-quadrupole interactions between nucleons have been reported, which may give rise to a variety of collective effects \cite{Trainor}. The quadrupole-quadrupole interactions between nuclei and impurity paramagnetic ions in metals and alloys were indirectly analyzed through conduction electrons \cite{Fazleyev}. It has been shown that the quadrupole-quadrupole interactions play a crucial role in driving nuclear-excited states \cite{Escher}. 
Moreover, density broadening has been studied in different disciplines. The quadrupolar interactions between two ions have been examined \cite{Morin}. Quadrupole-quadrupole interactions have been used to study plasma-induced transparency \cite{Rana}. Furthermore, quadrupole-quadrupole interactions contribute to the C$_5$ coefficient, which has been calculated for Cs \cite{Lepers}. Quadrupole-quadrupole interactions have been studied in molecules \cite{Liu, Chaback}. The angular dependence of quadrupole-quadrupole interactions is investigated in ultracold gases \cite{Wang}. %The density broadening stems from the different orders of multipole-multipole interactions. One of the very common multipole-multipole interactions is the quadrupole-quadrupole interactions. Quadrupole-quadrupole interactions have been studied \cite{}. 
However, most of the studies focus on two-body one-dimensional interactions. In this article, we focus on the three-dimensional few-body quadrupole-quadrupole interactions, and lower-order interactions, such as dipole-dipole interactions and dipole-quadrupole interactions among neutral atoms, are ignored. The on-resonance dipole-dipole interactions, which are proportional to $\frac{1}{R^3}$, are negligible if all the atoms are in one state. The van der Waals interactions, which are proportional to $\frac{1}{R^6}$, are more likely to cause asymmetric broadening. On-resonance quadrupole-quadrupole interactions, which are proportional to $\frac{1}{R^5}$, are experimentally tested \cite{Jianing} in laser-cooled \cite{Wineland, Neuhauser, William, Chu, Lett, Ketterle, Cornell} highly excited states \cite{Gallagher4, Dunning}. Such interactions have a variety of applications, such as quadrupole-quadrupole blockade for quantum gates \cite{Jaksch, Lukin, Saffman, Walker, Pierre, Alex, Ryabtsev, Li, Gunther}, quantum information storage at higher densities, and quadrupole-quadrupole coupled molecules. 

This paper is arranged in the following way: the quadrupole-quadrupole interactions are presented in the next section, which is
followed by the discussions about applications to different cases.

\section{Theory}
\begin{figure}[tpb]
\centering
\includegraphics[width=8cm,angle= 0 ,height=8cm]{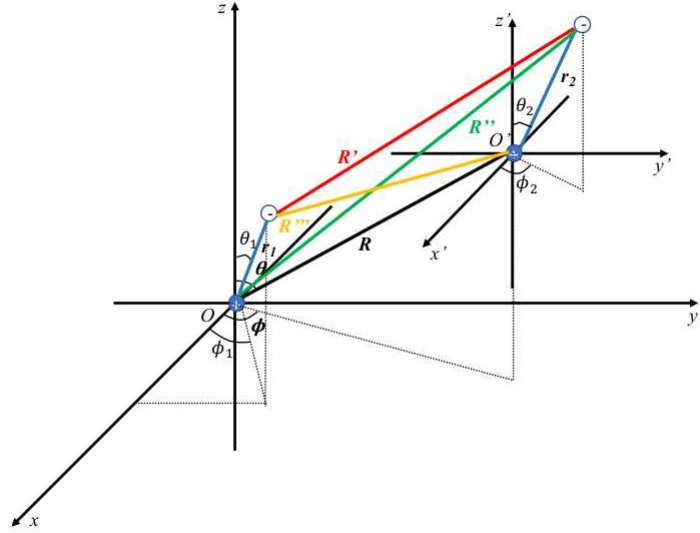}
\caption{Two ion cores of the two atoms, atom 1 (the bottom atom) and atom 2 (the top atom), are located at the origins of the two Cartesian coordinates, $O$ and $O'$. The distance between the bottom ion core and the electron of the bottom atom is $r_1$, and the distance between the top ion core and the electron of the top atom is $r_2$. The angle between $r_1$ and the $z$ axis is $\theta_1$, and the angle between the projection of $r_1$ in the $x-y$ plane and the $x$ axis is $\phi_1$. The angle between $r_2$ and the $z'$ axis is $\theta_2$, and the angle between the projection of the $r_2$ in the $x'-y'$ plane and the $x'$ axis is $\phi_2$. The distance between the two ion cores is $R$, the distance between the two electrons is $R'$, the distance between the ion core of the bottom atom and the electron of the top atom is $R''$, and the distance between the electron of the bottom atom and the ion core of the top atom is $R'''$. } 
\label{Three_D_schematic}
\end{figure}

Quadrupole-quadrupole interactions can be calculated by expanding the Coulomb interactions between four charges \cite{Jackson}. In this article, we assume the four charges are from two atoms separated by a distance $R$ as shown in Fig.  \ref{Three_D_schematic}. We further assume that the internuclear spacing, $R$, is much greater than the size of the atoms, $r_1$ and $r_2$. The Born-Oppenheimer approximation is applied. Specifically, the positive ions have zero kinetic energy. The Hamiltonian for this system is
\begin{equation}
\begin{split}
H=-\frac{\hslash}{2\mu _1}\triangledown _1-\frac{\hslash}{2\mu _2}\triangledown _2+V_1+V_2+V_{12}
 \end{split}
\label{H}
\end{equation}
where $\mu_1$ and the $\mu_2$ are the effective mass of atom 1 and 2, or the bottom atom and the top atom. The first two terms are the kinetic energy of the two atoms. The third and fourth terms are the internal Coulomb potential energy within atom 1 and atom 2 respectively, $V_1=-\frac{e^2}{4\pi \epsilon_0 r_1}$ and $V_2=-\frac{e^2}{4\pi \epsilon_0 r_2}$. $V_{12}$ is the quadrupole-quadrupole interactions between four charges shown in Fig. \ref{Three_D_schematic}, which can be written as: 
\begin{equation}
\begin{split}
 V_{12}&=\frac{Q_1Q_2}{4\pi \epsilon_0R^5}\{
 \frac{35}{8}sin^4\theta(e^{4i\phi}C_{2,-2}^{(1)}C_{2,-2}^{(2)}+e^{-4i\phi}C_{2,2}^{(1)}C_{2,2}^{(2)})\\&+\frac{35}{8}sin^2\theta sin(2\theta)[e^{3i\phi}(C_{2,-2}^{(1)}C_{2,-1}^{(2)}+C_{2,-1}^{(1)}C_{2,-2}^{(2)})-e^{-3i\phi}(C_{2,2}^{(1)}C_{2,1}^{(2)}+C_{2,1}^{(1)}C_{2,2}^{(2)})]\\&
 +(7cos^2\theta-1)\{\frac{5}{8}\sqrt{6}sin^2\theta[e^{-2i\phi}(C_{2,2}^{(1)}C_{2,0}^{(2)}+C_{2,0}^{(1)}C_{2,2}^{(2)})+e^{2i\phi}(C_{2,-2}^{(1)}C_{2,0}^{(2)}+C_{2,0}^{(1)}C_{2,-2}^{(2)})]\\&+\frac{5}{2}sin^2\theta(e^{-2i\phi}C_{2,1}^{(1)}C_{2,1}^{(2)}+e^{2i\phi}C_{2,-1}^{(1)}C_{2,-1}^{(2)})\}
 \\&+(\frac{7}{4}sin^2\theta -1)\frac{5}{2}sin(2\theta)[e^{-i\phi}(C_{2,2}^{(1)}C_{2,-1}^{(2)}+C_{2,-1}^{(1)}C_{2,2}^{(2)})-e^{i\phi}(C_{2,-2}^{(1)}C_{2,1}^{(2)}+C_{2,1}^{(1)}C_{2,-2}^{(2)})]\\&
 +(3-7cos^2\theta)\frac{5}{8}\sqrt{6}sin(2\theta)[e^{-i\phi}(C_{2,0}^{(1)}C_{2,1}^{(2)}+C_{2,1}^{(1)}C_{2,0}^{(2)})-e^{i\phi}(C_{2,0}^{(1)}C_{2,-1}^{(2)}+C_{2,-1}^{(1)}C_{2,0}^{(2)})]\\&
 +(\frac{35}{8}sin^4\theta -5sin^2\theta +1)(C_{2,-2}^{(1)}C_{2,2}^{(2)}+C_{2,2}^{(1)}C_{2,-2}^{(2)})+(\frac{105}{4}cos^4\theta -\frac{45}{2}cos^2\theta +\frac{9}{4})C_{2,0}^{(1)}C_{2,0}^{(2)}\\&
 +[-\frac{35}{8}sin^2(2\theta )+\frac{5}{2}cos^2\theta +\frac{3}{2}](C_{2,-1}^{(1)}C_{2,1}^{(2)}+C_{2,1}^{(1)}C_{2,-1}^{(2)})\}.
\end{split}
\label{Vdd3d}
\end{equation}
%The first term and second term in Eq. \eqref{H} represent the kinetic energy for the bottom electron, electron 1, and the upper electron, electron 2, respectively. 
where $Q_1=er_1^2$ and $Q_2=er_2^2$ are the quadrupole moments of atom 1 and atom 2. $R$ is the distance between the two atoms as shown in Fig. \ref{Three_D_schematic}. $C_{2,q}$ is the second-order spherical tensor in Edmonds \cite{Edmonds}. Here $q=-2, -1, 0, 1, 2$. It is shown that if $\theta =0$ and $\phi =0$, Eq. \eqref{Vdd3d} reduces to the one-dimensional quadrupole-quadrupole interactions \cite{Wang}. In addition, it can be proved that by reversing the R direction shown in Fig. \ref{Three_D_schematic}, or changing $\theta$ to $180^o-\theta$ and $\phi$ to $\phi +180^o$, the $V_{12}$ remains the same. Unlike the one-dimensional quadrupole-quadrupole interactions, the M, the projection of the total angular momentum of all three atoms in the z-axis, is no longer conserved in the three-dimensional quadrupole-quadrupole interactions. Therefore, the number of matrix elements is significantly increased for the three-dimensional quadrupole-quadrupole interactions compared to the one-dimensional quadrupole-quadrupole interactions. 

For three-body quadrupole-quadrupole interactions, the Hamiltonian can be modified from Eq. \eqref{H} by adding an additional atom:
\begin{equation}
\begin{split}
H=-\frac{\hslash}{2\mu _1}\triangledown _1-\frac{\hslash}{2\mu _2}\triangledown _2-\frac{\hslash}{2\mu _3}\triangledown _3+V_1+V_2+V_3+V_{qq},
 \end{split}
\label{H2}
\end{equation}
where $-\frac{\hslash}{2\mu _3}\triangledown _3$ and $V_3=-\frac{e^2}{4\pi \epsilon_0 r_3}$ are the kinetic and internal potential energy of atom 3. $V_{qq}$ is the sum of all pair-wise potential energies between the three atoms:
\begin{equation}
\begin{split}
V_{qq}=V_{12}+V_{23}+V_{31}.
 \end{split}
\label{H2}
\end{equation}
Very similar to the quadrupole-quadrupole interactions between atom 1 and atom 2 shown in Eq. \eqref{Vdd3d}, $V_{23}$ is the quadrupole-quadrupole interaction potential energy between atom 2 and atom 3, and $V_{31}$ is the quadrupole-quadrupole interaction between atom 3 and atom 1.

\section{Results and discussion}
\begin{figure}[tpb]
\centering
\includegraphics[width=5cm,angle= 0 ,height=5cm]{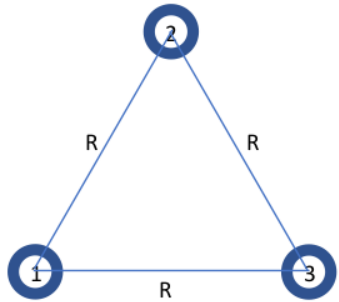}
\caption{  Three atoms are located at the three vertices of an equilateral triangle, and each atom is composed of one ion core and one electron. The distance between two neighboring atoms is R. } 
\label{Schematic_plot}
\end{figure}

In this article, we consider two-dimensional three-atom interactions. Specifically, the ddd Rydberg states \cite{Gallagher4, Dunning} are considered. The initial state is 34d$_{5/2,1/2}$34d$_{5/2,1/2}$34d$_{5/2,1/2}$. In other words, three 34d$_{5/2,1/2}$ atoms are located at the three vertices of an equilateral triangle as shown in Fig. \ref{Schematic_plot}. The total number of elements considered is (2+10)$^3$=12$^3$=1728, since the number of m states in the s state is 2, and the total number m states in the d states is 10 (4 in the d$_{3/2}$ states and 6 in the d$_{5/2}$ states), where m is the projection of the total angular momentum of one atom in the z-axis. The exponential part, 3, comes from the fact that three atoms are considered.

\begin{figure}[tpb]
\centering
\includegraphics[width=12cm,angle= 0 ,height=12cm]{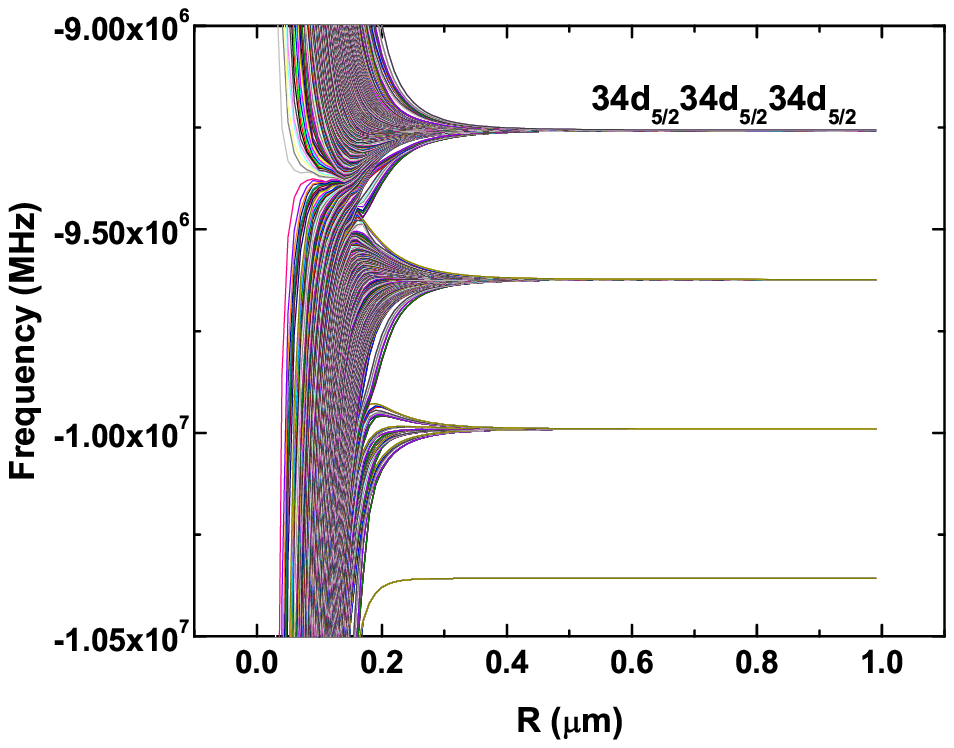}
\caption{  The energy, or frequency, as a function of the internuclear spacing. The energy levels calculated for the 34d34d34d, and the initial state is 34d$_{5/2,1/2}$34d$_{5/2,1/2}$34d$_{5/2,1/2}$. } 
\label{energy_levels}
\end{figure}

In the right-hand side of Fig. \ref{energy_levels}, the top energy level is 34d$_{5/2}$34d$_{5/2}$34d$_{5/2}$ at R=1 $\mu$m. The second highest levels are 34d$_{3/2}$34d$_{5/2}$34d$_{5/2}$, 34d$_{5/2}$34d$_{3/2}$34d$_{5/2}$, and 34d$_{5/2}$34d$_{5/2}$34d$_{3/2}$, and all three levels are degenerate. The third levels are 34d$_{3/2}$34d$_{3/2}$34d$_{5/2}$, 34d$_{5/2}$34d$_{3/2}$34d$_{3/2}$, and 34d$_{3/2}$34d$_{5/2}$34d$_{3/2}$, and all three are degenerate. The bottom level is 34d$_{3/2}$34d$_{3/2}$34d$_{3/2}$ at R=1 $\mu$m. Similar to other calculations, the energy levels are closely packed at shorter internuclear spacings, and the level structure is sophisticated. However, there are some general features. For example, the bottom level shows purely attractive quadrupole-quadrupole interactions, and some levels show both attractive and repulsive quadrupole-quadrupole interactions. There are still some open areas, which will be transparent for certain frequencies. Such a structure is very useful for creating transparent materials at a certain frequency range and block other frequencies. It is also shown that potential wells can be created through quadrupole-quadrupole interactions in a two-dimensional case. Moreover, potential peaks are observed. Those peaks can be as high as 100 THz or higher, which can possibly be used for quantum information storage. 

\begin{figure}[tpb]
\centering
\includegraphics[width=12cm,angle= 0 ,height=12cm]{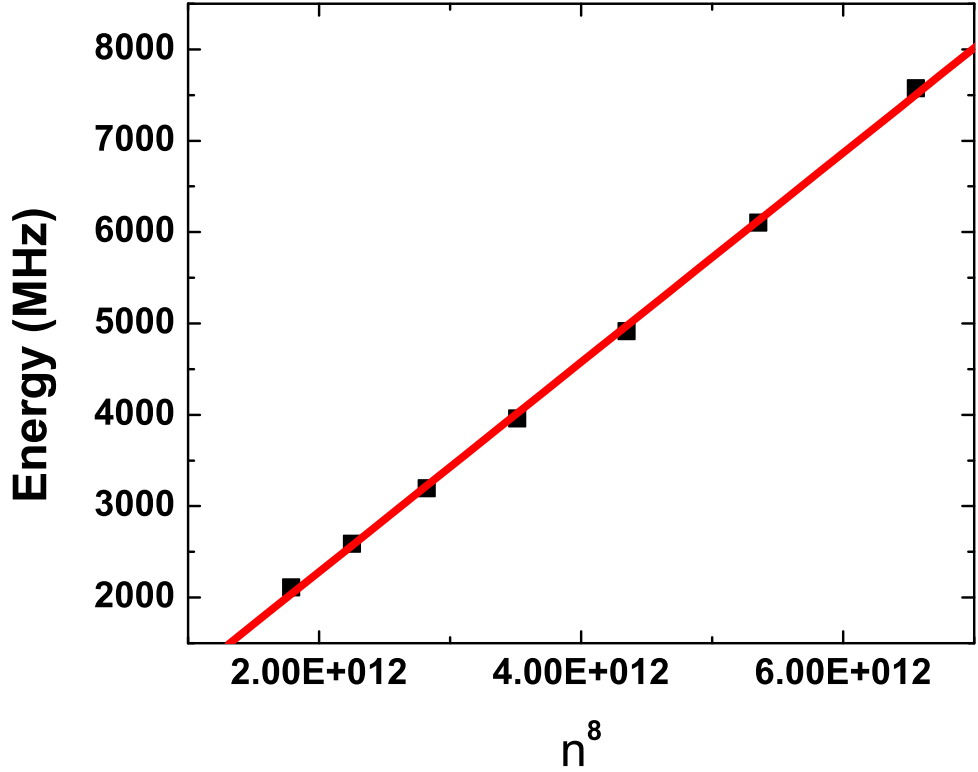}
\caption{  The maximum frequency shift in the higher frequency end at R=0.5 $\mu$m ($\blacksquare$) as a function of n$^8$, where n is the principal quantum number. The red line (\textcolor{red}{---}) is the linear fit of the calculated data.} 
\label{n_dependence}
\end{figure}

Fig. \ref{n_dependence} shows the principal quantum number n dependence of the maximum frequency shift. Specifically, the maximum repulsive frequency shift at the internuclear spacing R=0.5 $\mu$m vs. $n^8$ is plotted. It is shown that the linear fit is very close. The reason is the radius of the atoms is proportional to $n^2$, and each quadrupole moment has $r^2$ in it. Therefore, it is expected that the total shift is proportional to $n^8$. %However, it was observed that the maximum frequency shift vs. $n^9$ has an even better linear fit, which might be caused by the energy spacing between the coupled levels. 

\section{Conclusion}
In summary, the three-dimensional quadrupole-quadrupole interaction potential energy is expressed. In addition, this expression is applied to the three atoms located at three vertices of an equilateral triangle, and all three atoms are in the d states. Moreover, the principal quantum number n dependence is investigated. It is shown that the maximum repulsive frequency shift depends on $n^8$.

\section{Acknowledgement}  
    
It is a pleasure to acknowledge the support from the Air Force Office of Scientific Research (AFOSR), Army Research Office (USA), the University of South Alabama Faculty Development Council (USAFDC).

\end{document}